\documentclass[preprint,12pt]{elsarticle}

\usepackage{amssymb}

\journal{Solid State Communications}

\begin{document}

\begin{frontmatter}



\title{Electronic configuration of Mn ions in the $\pi$-$d$ molecular ferromagnet $\beta$-Mn phthalocyanine studied by soft x-ray magnetic circular dichroism}


\author{T. Kataoka, Y. Sakamoto, Y. Yamazaki, and V. R. Singh}

\address{Department of Physics and Department of Complexity Science and Engineering, 
University of Tokyo, Bunkyo-ku, Tokyo 113-0033, Japan}

\author{A. Fujimori}
\address{
Department of Physics and Department of Complexity Science and Engineering, University of Tokyo, Bunkyo-ku, Tokyo 113-0033, Japan;\\ 
Synchrotron Radiation Research Unit, Japan Atomic Energy Agency, Sayo-gun, Hyogo 679-5148, Japan
}

\author{Y. Takeda, T. Ohkochi, S.-I. Fujimori, T. Okane and Y. Saitoh}
\address{
Synchrotron Radiation Research Unit, Japan Atomic Energy Agency, Sayo-gun, Hyogo 679-5148, Japan
}

\author{H. Yamagami}
\address{
Synchrotron Radiation Research Unit, Japan Atomic Energy Agency, Sayo-gun, Hyogo 679-5148, Japan;\\
Department of Physics, Faculty of Science, Kyoto Sangyo University, Kyoto 603-8555, Japan
}

\author{A.~Tanaka}
\address{Department of Quantum Matter, ADSM, Hiroshima University, Higashi-Hiroshima 739-8530, Japan
}

\begin{abstract}
We have studied the electronic structure of the molecular ferromagnet 
$\beta$-Mn phthalocyanine ($\beta$-MnPc) in a polycrystalline form, 
which has been reported to show ferromagnetism at T$<$8.6 K, 
by x-ray absorption spectroscopy (XAS) and x-ray magnetic circular dichroism (XMCD). 
From the experimental results and subsequent cluster-model calculation, 
we find that the ferromagnetic Mn ion in $\beta$-MnPc is largely in the $^4$$E$$_g$ ground state arising from the 
($e$$_{g}$)$^3$($b$$_{2g}$)$^1$($a$$_{1g}$)$^1$ [($d_{xz,yz}$)$^3$($d_{xy}$)$^1$($d_{z^{2}}$)$^1$] 
configuration of the Mn$^{2+}$ state. 
Considering that the highest occupied molecular orbital (HOMO) of MnPc with the $^4$$E$$_g$ ground 
state originates from the $a$$_{1g}$ orbital of the Mn$^{2+}$ ion, it is proposed that $a$$_{1g}$-$a$$_{1g}$ exchange coupling via the 
$\pi$ orbitals of the phthalocyanine ring plays a crucial role in the ferromagnetism of $\beta$-MnPc.
\end{abstract}


\end{frontmatter}


\newpage
\section{Introduction}
\label{}
Transition-metal phthalocyanine (Pc) molecules [see Fig. 1(a)] have attracted great interest in materials science because of 
their wide applications in the areas of gas-sensing devices, biological engineering, 
organic field effect transistors and pigments \cite{Leznoff, Carter, Kahn, Tu, Forrest, Mattioli}. 
Their interesting magnetic properties, e.g., ferromagnetism and manipulating magnetism, 
of the transition-metal Pc \cite{Chen, Fu, Crommie, Heutz, Wu} 
allow us to design the new generation molecular electronics with enhanced functionalities such as the molecular ferromagnets. 
In particular, there has been interest in the ferromagnetism of the $\beta$ polymorph of Mn phthalocyanine ($\beta$-MnPc) \cite{Barraclough, Lever, Miyoshi, Awaga, Yamada}. 

The magnetic susceptibility of $\beta$-MnPc obeys the Curie-Weiss law with a positive Curie-Weiss constant 
and the Curie temperature $T_c$ is determined to be 8.6 K \cite{Barraclough, Lever, Miyoshi}. 
In $\beta$-MnPc, each Mn atom is located directly above or below the N atoms of the adjacent 
parallel MnPc molecules, as shown in Fig. 1(b), and the distance 3.38 $\rm{\AA}$ between the Mn and adjacent N atoms 
is shorter than that in other transition-metal Pc's in the $\beta$-type crystalline form \cite{Yamada}. 
Also, ferromagnetism has been observed experimentally only in the $\beta$-type crystalline form \cite{Barraclough, Lever, Miyoshi, Awaga, Yamada}. 
Therefore, it has been expected that the ferromagnetism comes from super-exchange interaction 
between nearest-neighboring Mn atoms via the $\pi$ orbitals of 
the Pc ring, namely, $\pi$-orbital-mediated intermolecular interaction \cite{Barraclough, Awaga, Yamada}. 
On the other hand, according to recent density functional theory (DFT) calculations \cite{Heutz, Wu}, 
it has been pointed out that the super-exchange interaction should lead to structure-independent antiferromagnetic correlations. 
Thus, magnetic coupling in $\beta$-MnPc has not been clarified yet and 
more information about the electronic configuration of Mn ions in $\beta$-MnPc is necessary to elucidate the nature of magnetic interaction.

The molecular symmetry of MnPc is almost square planar coordination ($D_{4h}$), with tiny distortions out of the molecular plane \cite{Mason, Liao, Kahl}. 
In the $D_{4h}$ point group symmetry of MnPc, 
the Mn 3$d$ level is split into $a_{1g}$($d_{z^{2}}$), $b_{1g}$($d_{x^{2}-y^{2}}$), $b_{2g}$($d_{xy}$), and $e_{g}$($d_{xz,yz}$). 
Magnetic circular dichroism (MCD) and UV-visible measurements of MnPc \cite{Williamson} suggest 
that the ground state of the Mn$^{2+}$ ion is a $^4$$E$$_g$ state with the ($e_g$)$^3$($b_{2g}$)$^1$($a_{1g}$)$^1$ configuration, as shown in Fig. 1(c). 
On the other hand, the magnetic susceptibility and magnetization data at higher temperatures \cite{Barraclough} 
are only weakly anisotropic, suggesting a quartet spin state in which some excited $^4$$E$$_g$ 
term is mixed via spin-orbit coupling into the $^4$$A$$_{2g}$ one 
[($b_{2g}$)$^2$($e_g$)$^2$($a_{1g}$)$^1$, see Fig. 1(c)]. 
Because there has been little experimental information about the electronic configuration of Mn ions in $\beta$-MnPc nor 
the exchange interaction between MnPc molecules, 
clarifying the electronic structure of $\beta$-MnPc experimentally is expected 
to open up opportunities to develop new materials with novel functionalities 
in the field of the molecular ferromagnets. 

X-ray absorption spectroscopy (XAS) and x-ray magnetic circular dichroism (XMCD) in core-level absorption 
enable us to study the element specific electronic structure of $\beta$-MnPc. 
The line shapes of absorption and dichroism spectra are fingerprints of electronic structures 
such as the valence of Mn ions and the symmetry of crystal fields. 
In particular, XMCD is a powerful tool to study element-specific local magnetic states \cite{Funkb}. 
Here, we report on the magnetic-field dependence of dichroism at the Mn 2$p$$\rightarrow$3$d$ absorption edge to study 
the local electronic structure of magnetically active Mn ions. 
We also discuss the ground state of the ferromagnetic (FM) Mn ion in $\beta$-MnPc based on the experimental results and 
subsequent cluster-model calculation \cite{Tanaka1, Tanaka2, Mizokawa}. 
Since the Mn ion in MnPc is coordinated by N atoms and can be 
described by the MnN$_4$ cluster model, the cluster-model 
calculation provides us with information about the ground state of the Mn ion \cite{Mizokawa, AEBouq, Eskes}.

\section{Experimental}
\label{}

The crystal structure of $\beta$-MnPc is body centered monoclinic with 
$a$ = 20.2 $\rm{\AA}$, $b$ = 4.75 $\rm{\AA}$, $c$ = 15.1 $\rm{\AA}$, 
and $\beta$ = 121.7$^\circ$ \cite{Miyoshi}. 
Single crystals of $\beta$-MnPc were obtained from MnPc powders (Aldrich) 
purified by the vacuum sublimation ($\sim$ 5$\times$10$^{-3}$ Pa) and we confirmed 
the purity of the material by the mass spectrum of our samples using 
the matrix-assisted laser desorption ionization time-of-flight mass spectrometry (MALDITOFMS) method. 
A large number of single crystals of $\beta$-MnPc were powdered as found by an earlier study \cite{Miyoshi}. 
The magnetization curve measured with a superconducting quantum interference device (SQUID) 
indicated that $\beta$-MnPc in a polycrystalline form exhibited ferromagnetism below $T$ $\sim$ 8.6 K as shown in Fig. 1(d), 
consistent with the earlier magnetization measurements of $\beta$-MnPc \cite{Barraclough}. 
These results suggest that our samples are $\beta$-type crystalline form. 
Assour and Kahn \cite{Asson} have reported that the $\beta$-phthalocyanine is invariably the only polymorph 
prepared by the sublimation methods, which are found to be the most effective purification and 
crystal growth processes for these metallo-organic complexes. 
For the XAS and XMCD measurements, we put powders consisting of $\beta$-MnPc single crystals on 
the copper sample holder. We note that XMCD is an element-specific tool and the Mn 2$p$$\rightarrow$3$d$ XMCD 
signals are free from possible magnetic signals from the sample holder.

XAS and XMCD measurements were performed at the undulator beam line BL-23SU of SPring-8 \cite{YSaitoh,JOkamoto} and 
spectra were recorded in the total-electron-yield (TEY) mode (probing depth$\sim$5 nm). 
The degree of circular polarization was higher than $\sim$ 95\%. 
The monochromator resolution was $E$$/$$\Delta$$E$ $>$ 10000. 
Magnetic fields $H$ up to 8 T were applied perpendicular to the sample surface. 
All the experiments were performed at $T$$\sim$6 K (read temperature). 
However, due to thermal contact, the actual sample temperature is at most $\sim$ 1 K higher than read temperature using a 
thermocouple attached to the sample holder.

\section{Results and Discussion}
\label{}

Figure 2 shows the Mn 2$p$$\rightarrow$3$d$ XAS and XMCD spectra of $\beta$-MnPc taken at $H$$=$8 T. 
Here, $\mu$$^+$ and $\mu$$^-$ refer to absorption spectra for photon helicity parallel and 
antiparallel to the Mn 3$d$ spin, respectively. 
The structures around $h\nu$ $=$ 642 and 653 eV are due to absorption from 
the $j$ $=$ 3/2 ($L_3$) and 1/2 ($L_2$) components of the Mn 2$p$ core levels, respectively. 

Figures 3(a) and (b) show the Mn 2$p$$\rightarrow$3$d$ XMCD spectra of $\beta$-MnPc measured at various magnetic fields. 
In Fig. 3(c), we have evaluated the orbital magnetic moment ($M_\mathrm{orb}$), 
the spin magnetic moment ($M_\mathrm{spin}$), and the total magnetic moment 
($M_\mathrm{tot.}$=$M_\mathrm{spin}$ + $M_\mathrm{orb.}$) of the Mn$^{2+}$ (d$^5$) ion 
as functions of magnetic field using the integrals $p$, $q$, and $r$ indicated in Fig. 2 \cite{Thole,Garra}:

\begin{equation}
\label{eqn:XMCDorb}
M_\mathrm{orb} = - \frac{4 (\Delta A_{L_3} + \Delta A_{L_2})}
                         {3 (A_{L_3} + A_{L_2})}
                            n_h\mu_B,
\end{equation}
\begin{equation}
\label{eqn:XMCDspin}
M_\mathrm{spin} + 7M_\mathrm{T} = 
 - \frac{2 (\Delta A_{L_3} - 2\Delta A_{L_2})}
                         {A_{L_3} + A_{L_2}}
                            n_h\mu_B,
\end{equation}
$\noindent$
where $A_{L_2}$ ($A_{L_3}$), and $\Delta A_{L_2}$ ($\Delta A_{L_3}$) are the $L_2$- ($L_3$-) 
edge integrated XAS and XMCD intensities, respectively, $n_h$ 
is the 3$d$ hole number, and $M_\mathrm{T}$ is the magnetic quadrapole moment. 
Here, we assumed $n_h$ to be 5 because the formal number of 3$d$ electrons of Mn$^{2+}$ ions in MnPc is 5. 
In general, the value of $M_\mathrm{T}$ is negligibly small compared to spin and orbital magnetic moments. 
Therefore, we assumed $M_\mathrm{T}$ to be zero, following the earlier works \cite{Yamamoto,Mamiya}. 
Due to the relatively small $L$$_2$-$L$$_3$ spin-orbit splitting 
compared to the 2$p$-3$d$ Coulomb-exchange interaction, the spin magnetic moments estimated from XMCD is underestimated. 
According to an atomic calculation  \cite{Teramura} taking into account the multipole electron-electron interaction, 
the deviation estimated from XMCD results is about 32\% for Mn$^{2+}$. 
Therefore, we considered the correction factor \cite{Teramura} in the estimation of the spin magnetic moments from XMCD results. 
Figure 3(c) shows almost paramagnetic behavior, but one observes finite XMCD signals down to $H$$=$0.1 T as shown in Figs. 3(a) and 3(b), 
indicating the presence of FM component due to the Mn ionic states. 
The observed paramagnetic XMCD signals may have originated from the presence of different crystalline phases or impurities on the powder sample. 
We consider that the magnetic-field-dependent XMCD signals in this sample mainly come from 
isolated and uncoupled MnPc molecules and/or antiferromagnetically coupled canted magnetic moments in the $\alpha$-phase. 
Indeed, the negative Curie–Weiss constant, indicating the existence of the antiferromagnetic interaction, 
was observed in $\alpha$-phase films of MnPc \cite{Heutz}.

In Fig. 4(a), the Mn 2$p$$\rightarrow$3$d$ XMCD spectrum at $H$$=$0.1 T is compared with cluster-model calculations for the 
$^4$$E$$_g$ and $^4$$A$$_{2g}$ ground states of the Mn$^{2+}$($d^5$) ion. 
The XMCD spectrum at $H$$=$0.1 T is expected to be mostly due to the FM component, 
because the paramagnetic XMCD signals is proportional to magnetic field. 
In the planar square MnN$_4$ cluster with the $D_{4h}$ symmetry, 
the Mn 3$d$-N 2$p$ transfer integrals $T(\gamma)$ can be described using 
the Slater-Koster parameters $(pd\sigma)$ and $(pd\pi)$: $T(b_{1g})=\sqrt{3}(pd\sigma)$, 
$T(a_{1g})=(pd\sigma)$, $T(e_g)$ $=\sqrt{2}(pd\pi)$ and $T(b_{2g})=2(pd\pi)$ \cite{FKotani,Harrison}. 
We have assumed that the Mn 3$d$ orbital energies $\varepsilon(\gamma)$ are given by $\varepsilon(\gamma)=\alpha T^2(\gamma)$. 
For the Jahn-Teller active $^4$$E$$_g$ configuration, a small $D_{2h}$ distortion of the MnPc molecule caused 
by the Jahn-Teller effect is further included by differentiating the values of hopping integrals for two 
kinds of $e_g$ orbitals (the $yz$ and $zx$ orbitals): 
$T(e_g)=\sqrt{2}(pd\pi)(1\pm\delta/2)$. 
Since the MnPc molecule is highly anisotropic and polycrystalline samples were used in the experiments, 
the XMCD spectra are averaged over all possible directions of the molecule in the calculations. 
The magnetization direction is assumed to be parallel to the magnetic field. 
Note that the spin-orbit interaction of the Mn 3$d$ orbitals has been taken into account in these calculations. 

The best fit to the Mn 2$p$$\rightarrow$3$d$ XMCD spectrum at $H$ $=$ 0.1 T of $\beta$-MnPc can be achieved with the electronic structure 
parameters $\alpha$ = 0.28~eV$^{-1}$, ($pd\sigma$) = -1.73~eV, ($pd\pi$) = 0.75~eV, $\delta=0.1$, 
the charge-transfer energy $\Delta=1.0$~eV, the averaged Coulomb energy between $3d$ 
electrons $U_{dd}=5.0$~eV, and the core-hole potential $U_{dc}=6.0$~eV. 
The charge-transfer energy $\Delta$ is defined as [E(3$d^{n+1}\underline L$)-E(3$d^n$)], 
where $E(3d^{n+1}\underline L$) and $E(3d^n$) are the average energies of 3$d^{n+1}\underline L$ and 3$d^n$ configurations, respectively. 
The peak positions and the relative intensities among the peaks are reproduced well with this parameter set. 
The Mn 3$d$ level ordering in the initial state is  $b_{1g}$$>>$$a_{1g}$$>$$b_{2g}$$>$$e_g$ 
and the ground state is the intermediate spin $S=3/2$ state with the $^4$$E$$_g$ configuration \cite{Barraclough,Liao}. 
Thus, we consider that the ferromagnetically active component of $\beta$-MnPc 
comes largely from the $^4$$E$$_g$ ground state of the Mn$^{2+}$ ion. 
For comparison, the spectra with the $^4$$A$$_{2g}$ ground state is also calculated. 
The parameters are identical to the above calculations except for the values 
of the hopping integrals and energy levels of the $e_g$ and $b_{2g}$ orbitals being 
interchanged so as to obtain Mn 3$d$ level ordering $b_{1g}$$>>$$a_{1g}$$>$$e_g$$>$$b_{2g}$, 
which favors the $^4$$A$$_{2g}$ configuration \cite{Barraclough}. 
The calculated XMCD spectrum of the $^4$$A$$_{2g}$ ground state does not reproduce the FM component of the experimental one. 
Other parameter sets have been also examined for the spectra with the $^4$$A$$_{2g}$ ground state. 
However, we found rather poor agreement with the experimental spectrum. 
We also found that the experimental XMCD spectrum cannot be explained with the high-spin ($^6$$A$$_{1g}$) ground state.

Based on the DFT\cite{Liao} and DV-X$\alpha$\cite{Suzuki} calculations, 
the highest occupied molecular orbital (HOMO) of MnPc with the $^4$$E$$_g$ ground state 
originates from the half-filled $a_{1g}$ orbital of the Mn ion. 
Since the HOMO of a molecule affects its physical properties, 
the $a_{1g}$ orbitals of Mn are expected to affect the ferromagnetism of $\beta$-MnPc. 
Considering the orbital symmetry of the $\pi$ orbitals of the Pc ring \cite{Barraclough, Yamada}, 
we suggest that Mn $a_{1g}$-$a_{1g}$ coupling via the $e_{g}$ orbitals of the Pc ring plays a crucial role in the 
FM coupling of $\beta$-MnPc, as shown in Fig. 4(b).

\section{Conclusion}
\label{}
We have performed Mn 2$p$$\rightarrow$3$d$ XMCD measurements on $\beta$-MnPc. 
From the magnetic-field dependence of dichroism at the Mn 2$p$$\rightarrow$3$d$ absorption edge, 
the XMCD signals due to the ferromagnetic Mn ions were observed. 
Based on the Mn 2$p$$\rightarrow$3$d$ XMCD results and subsequent cluster-model calculation, 
we find that the ferromagnetic Mn ion in $\beta$-MnPc is largely in the $^4$$E$$_g$ ground state arising from the 
($e_{g}$)$^3$($b_{2g}$)$^1$($a_{1g}$)$^1$ [($d_{xz,yz}$)$^3$($d_{xy}$)$^1$($d_{z^{2}}$)$^1$] configuration. 
Considering that the highest occupied molecular orbital (HOMO) of MnPc with the $^4$$E$$_g$ ground state 
originates from the $a$$_{1g}$ orbital of the Mn$^{2+}$ ion, it is proposed that $a$$_{1g}$-$a$$_{1g}$ exchange coupling via the 
$\pi$ orbitals of the phthalocyanine ring plays a crucial role in the ferromagnetism of $\beta$-MnPc. 

\section{Acknowledgments}
\label{}

We thank S. Gohda for providing us $\beta$-MnPc samples purified by sublimation, 
and H. Okamoto for his valuable technical support for the SQUID measurement. 
We also thank the Material Design and Characterization Laboratory, 
Institute for Solid State Physics, University of Tokyo, for the use of the SQUID magnetometer. 
The experiment at SPring-8 was performed under the approval of the Japan 
Synchrotron Radiation Research Institute (JASRI) (proposal no. 2008A3824). 
This work was supported by a Grant-in-Aid for Scientific Research (S22224005) from JSPS, Japan.





\bibliographystyle{model1a-num-names}
\bibliography{<your-bib-database>}



\section*{Figures}
\end{center}

\begin{figure}[htbp]
 \begin{center}
  \includegraphics[width=\linewidth]{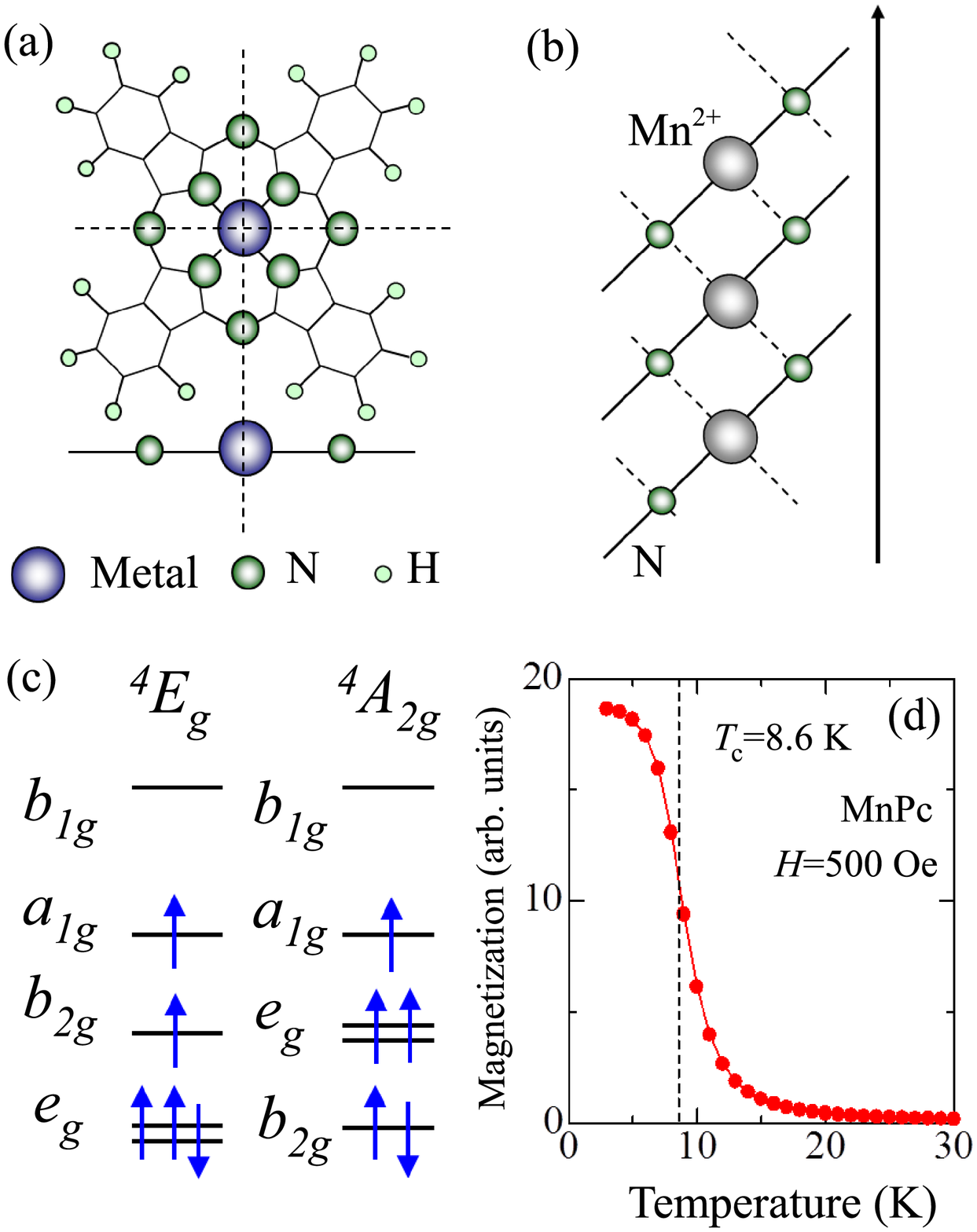}
 \end{center}
\caption{(Color online) Molecular structure and magnetic properties of 
manganese phthalocyanine (MnPc). 
(a) Molecular structure of the transition-metal phthalocyanine. 
Unlabeled atoms are carbon. 
(b) Molecular stacking structure along the $b$ axis of $\beta$-MnPc. 
(c) Two energy-level schemes for MnPc with $S$ $=$ 3/2, namely, 
the $^4$$E$$_g$ and $^4$$A$$_{2g}$ ground states. 
(d) The temperature dependence of magnetization at $H$ $=$ 500 Oe for $\beta$-MnPc samples. 
The measurement was made in field cooled ($H$ $=$ 500 Oe) and the samples were measured during warm-up.}
\end{figure}

\begin{figure}[htbp]
 \begin{center}
  \includegraphics[width=\linewidth]{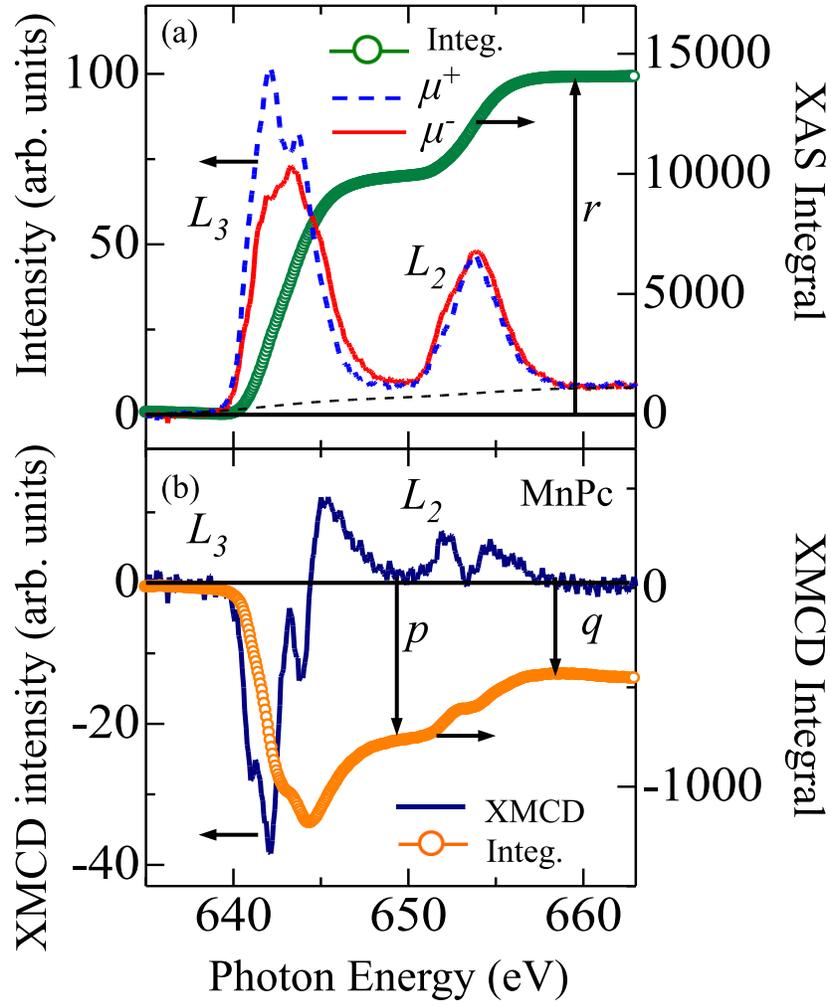}
 \end{center}
\caption{(Color online) 
Mn 2$p$$\rightarrow$3$d$ XAS (a) and XMCD (b) spectra of MnPc, 
recorded at $H$=8 T.}
\end{figure}

\begin{figure}[htbp]
 \begin{center}
  \includegraphics[width=\linewidth]{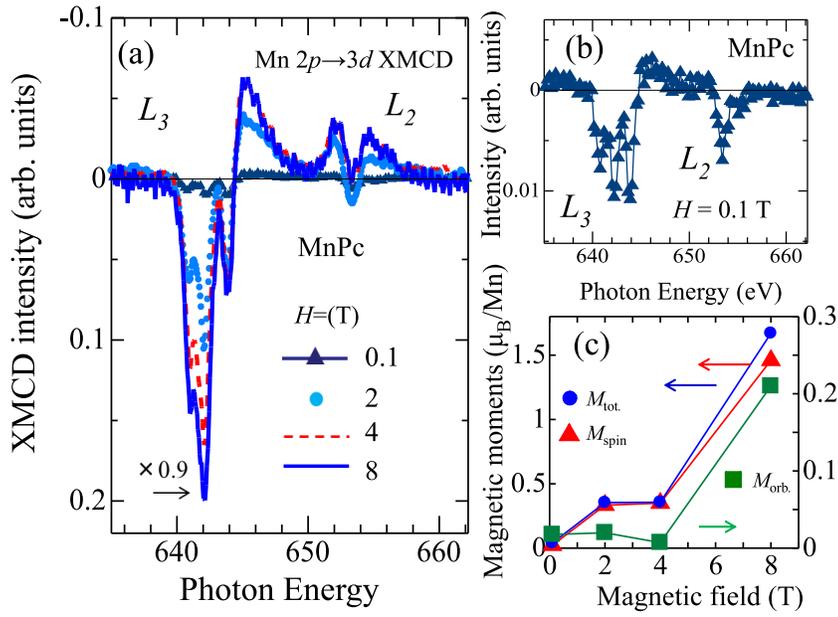}
 \end{center}
\caption{(Color online) 
Mn 2$p$$\rightarrow$3$d$ XAS and XMCD spectra of $\beta$-MnPc. 
(a) Mn 2$p$$\rightarrow$3$d$ XMCD spectra of $\beta$-MnPc measured at various magnetic fields. 
(b) A magnified view of the XMCD spectrum at $H$=0.1 T. 
(c) Magnetic moments at various magnetic fields obtained from the Mn 2$p$$\rightarrow$3$d$ XMCD spectra.}
\end{figure}

\begin{figure}[htbp]
 \begin{center}
  \includegraphics[width=\linewidth]{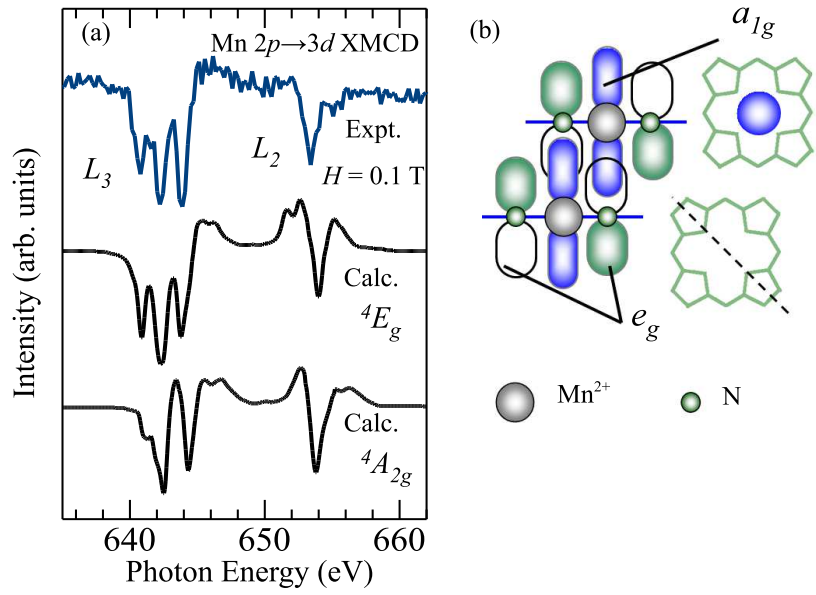}
 \end{center}
\caption{(Color online) 
Mn 2$p$$\rightarrow$3$d$ XMCD spectrum compared with cluster-model calculations. 
(a) Mn 2$p$$\rightarrow$3$d$ XMCD spectrum at $H$$=$0.1 T compared with calculations for the $^4$$E$$_g$ and $^4$$A$$_{2g}$ ground states. 
(b) Schematic picture of the $a$$_{1g}$-$a$$_{1g}$ coupling via the $e_{g}$ orbitals of 
the Pc ring in $\beta$-MnPc. The nodal plane (dashed line) of the $e_{g}$ orbital 
passes through two pyrrole N atoms and the central Mn$^{2+}$ ion.}
\end{figure}


\end{document}